Total 7 pages (including references)

# Empirical Evaluation of User Experience Using Lean Product and Process Development: A Public Institution Case Study in Indonesia

Aang Subiyakto[1, a)], Rohadatul Aisy[1, b)], Bernadus Gunawan Sudarsono[2, c)], Manorang Sihotang[3, d)], Didik Setiyadi[4, e)], Asrul Sani[5, f)]

[1]UIN Syarif Hidayatullah Jakarta, Jl. Juanda No. 95, Kota Tangerang Selatan 15412, Indonesia
[2]Universitas Bung Karno, Jl. Pegangsaan Timur No. 17A, Jakarta 10310, Indonesia
[3]Universitas Pramita Indonesia, Jl. Kampus Pramita, Tangerang 15810, Indonesia
[4]Universitas Bina Insani, Jl. Raya Siliwangi No. 6 Rawa Panjang, Bekasi 17114, Indonesia
[5]Asia e University, Jl. SS 15/4 No.106. 47500 Kuala Lumpur, Malaysia

[a)]Corresponding author: aang_subiyakto@uinjkt.ac.id, [b)]rohadatul.aisy16@mhs.uinjkt.ac.id,
[c)]gunawanbernadus@ubk.ac.id, [d)]manorang@raharja.com, [e)]didiksetiyadi@binainsani.ac.id,
[f)]asrul_sani09@yahoo.com

**Abstract.** The easiness, speed, accuracy, security are the four main indicators of information quality that may be of concern to people for accessing information in a public institution dashboard system. To find out whether the data displayed on the system is easily understood by the users, an empirical evaluation study may indispensable to be performed. Evaluating user experience is one way to know whether a system is in accordance with user needs or not, whether related to data, interfaces or system performance. This study assessed user experiences based on the product pyramid and its development process in the case of a public dashboard system in a public institution in Indonesia. The empirical experiment was conducted with about 15 participants using a scenario that were deliberately designed and then gave answers to the single ease question (SEQ) and the system usability scale (SUS). The results show that SEQ measurement results obtained value of almost 5.7 which indicates that the application measurement results have an easy level. The SUS results demonstrate that the system was in the acceptable category with the value of around 81.3. Although findings of the study might not contribute theoretically, the findings may demonstrate a practical consideration for the stakeholders, in terms of the next system development stage.

## INTRODUCTION

Transparency is an important issue for measuring the good governance performance of a government agency (1, 2). In Indonesia, this has been regulated in Law No. 14 the year of 2008 concerning Public Information Openness (3, 4). This law allows the public to obtain various information about policies and practices held by a government agency. Besides the information openness is one of the accountability forms as the institution that manages community fund, it is also a form of good communication management with stakeholders (5). In line with current developments in science and technology, the use of ICT is indisputably the compulsory implementation. In the country, various systems have been developed by government agencies and have absorbed much of the budget. However, performance evaluations of the use of ICTs in this information provision system tend to be carried out internally within the scope of project implementation. Independent testing related to this matter may still rarely done.

Besides, the high level of public attention to information access, the accuracy, speed, and accuracy of the data presented is the most important instrument in its presentation (6). To find out whether the data displayed on the system is easily understood and understood by the recipient of the information it is necessary to do an evaluation. Therefore, this study was conducted to find out whether a system is following user requirements or not, whether related to data, interfaces or system performance. Evaluation of user experience is one of the ways user experience is subjective and is focused on the use of a product, system, or service (7). In other words, user experience is how it feels to every user when the user is dealing directly with the system and using it. An information provider dashboard system in one of the government institutions is a case that is being investigated. The researchers used the lean product process pyramid method (8, 9). To guarantee the research performance, a question was then asked in this study about what is the easiness and acceptance levels the system use based on the user experience perceptions.

In the following parts, the second part of this paper presents the research methodology points. The third part explains the results and is discussion. This section describes sequentially the results of each evaluation phase and concludes with a discussion of the results based on the prior theoretical basis and research findings used in this study. This paper concludes with a conclusion part that concludes previous exposures related to research questions, limitations, and recommendations for further similar studies.

## RESEARCH METHODS

Procedurally, this evaluation study was carried out within six phases of the iterative process framework of the lean product pyramid (8, 9). Figure 1 demonstrates the phases. First, the researchers determined the target users of the system. Referring to the Law No. 14 the year of 2008 concerning Public Information Openness, the system was developed to provide information services to the public (3, 4). It is regarding implement good governance in the public institution. Therefore, the consumers or users of the system are citizen in the country. Second, the scholars identified underserved needs. This phase was conducted by studying literature about the quality constructs of public information. Here, the researchers adopted indicators of the information quality variable (Table 3). Third, the people defined the value proposition of the system by comparing the system with three similar dashboard systems to get a comparison value (Table 4). Fourth, the scholar reformulated features of the system to define the user needs. Here, use case diagram was used to design the features (Figure 2). Finally, the researchers evaluated the system using the user's experiences.

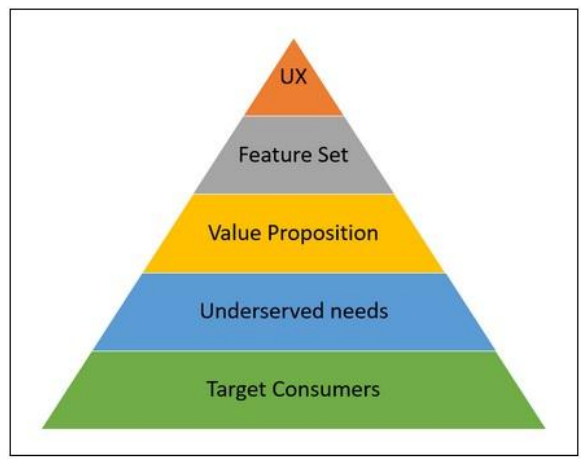

**FIGURE 1.** The Lean Product Development Pyramid (8, 9)

In detail, the user experience assessment was carried out within three sequential techniques, i.e., structured task scenario, single ease question (SEQ) testing (10, 11), and the system usability scale (SUS) (10, 12). SEQs were given into the participants after carrying out the task assigned to see the perception of the usefulness of the system.

SEQ consists of a question for each task with seven Likert's scale from very difficult to very easy (Table 1). The SUS test was to find out how high the level of usability and acceptability (acceptable) system design. SUS consists of 10 questions using five Likert's scale from disagreeing to agree (Table 2). Odd number questions (1, 3, 5, 7, and 9) are positive questions. While even number questions (2, 4, 6, 8, and 10) are negative questions. For this test, this involved five internal users from the internal institution that were the object of research and 10 people who were randomly selected as the participants. The assessment is based on three categories, namely Not Acceptable with a range of SUS scores 0-50.9, Marginal 51-70.9, and Acceptable 71-100.

**TABLE 1.** List of Task Scenario and SEQ Assessment

| No | Tasks | Assessment | | | | | | | | | |
|----|-------|------------|---|---|---|---|---|---|---|---|---|
| 1 | Display data of religious education! | Very Difficult | 1 | 2 | 3 | 4 | 5 | 6 | 7 | Very Easy |
| 2 | Display data of the Madrasah Tsanawiyah! | | 1 | 2 | 3 | 4 | 5 | 6 | 7 | |
| 3 | Display statistical data of the Raudhatul Athfal students! | | 1 | 2 | 3 | 4 | 5 | 6 | 7 | |
| 4 | Display data of Hajj and Umrah! | | 1 | 2 | 3 | 4 | 5 | 6 | 7 | |
| 5 | Display statistical data of Hajj data based on the participant's profession! | | 1 | 2 | 3 | 4 | 5 | 6 | 7 | |
| 6 | Display data of the management! | | 1 | 2 | 3 | 4 | 5 | 6 | 7 | |
| 7 | Display statistical data on the budget data based on its composition! | | 1 | 2 | 3 | 4 | 5 | 6 | 7 | |
| 8 | Display data on Islamic religion! | | 1 | 2 | 3 | 4 | 5 | 6 | 7 | |

**TABLE 2.** List of Usability Questions of SUS Assessment

| No | Usability Questions | Assessment | | | | | | |
|----|---------------------|------------|---|---|---|---|---|---|
| 1 | I seem to be using the system often. | Disagree | 1 | 2 | 3 | 4 | 5 | Agree |
| 2 | I see, there are parts of the system features that are quite troublesome, which should not need to happen. | | 1 | 2 | 3 | 4 | 5 | |
| 3 | I think, the system is easy to use. | | 1 | 2 | 3 | 4 | 5 | |
| 4 | I seem to need the help of a technician to be able to smoothly use the system. | | 1 | 2 | 3 | 4 | 5 | |
| 5 | I think the features of the system are already well integrated. | | 1 | 2 | 3 | 4 | 5 | |
| 6 | I found too many inconsistencies in the system. | | 1 | 2 | 3 | 4 | 5 | |
| 7 | I think people will be able to use the system very quickly. | | 1 | 2 | 3 | 4 | 5 | |
| 8 | I think the system is very difficult to use. | | 1 | 2 | 3 | 4 | 5 | |
| 9 | I feel great using this system. | | 1 | 2 | 3 | 4 | 5 | |
| 10 | I must learn many things first before starting to use the system. | | 1 | 2 | 3 | 4 | 5 | |

# RESULTS AND DISCUSSION

## The Target Users of the System

The phase determined the system market segmentation. The market segmentation of this system are people who need information that can be useful for them as public information that will be obtained in the form of public data. Here, the information is about educational information which manages by the public institution.

## The Identified Underserved User Needs

Table 3 presents results of the six identified user needs on the public dashboard system of the sampled institution, around the accuracy, clarity, easy to use, easy to understand, validity, and the newness issues of information.

**TABLE 3.** The Identified User Needs

| No | The User Needs |
|---|---|
| 1 | People need a dashboard system that displays information accurately. |
| 2 | People need a dashboard system that displays information. |
| 3 | People need a dashboard system that displays information is easy to use. |
| 4 | People need that information displayed on a dashboard system is easy to understand. |
| 5 | People need that information displayed on a dashboard system is valid to understand valid. |
| 6 | People need that information displayed on a dashboard system is always updated. |

## The Proposed Values

Table 4 shows the value proposition of the system by comparing the system with three similar dashboard systems, i.e., the National Statistics Office of Mongolia and the Philippine Statistic Authority.

**TABLE 4.** The Value Proposition of the Dashboard System

| Comparison Aspects | The sampled system | National Statistics Office of Mongolia | Philippine Statistic Authority |
|---|---|---|---|
| The attraction of the infographics | √ | √ | √ |
| Easiness of the information understanding | √ | √ | √ |
| Easiness achievement of the system | √ | √ | √ |
| The satisfaction of the information display | √ | √ | √ |
| Attractive interface design | √ | √ | √ |
| The attraction of the used colour combination | √ | √ | - |
| Easiness of the system use | √ | √ | √ |
| The usefulness of the system for all fields of life | - | - | √ |
| Easiness of the system menus to be operated | √ | √ | √ |

**TABLE 5.** Results of SEQ Assessment

| R | \multicolumn{8}{c}{Tasks} | | Σ |
|---|---|---|---|---|---|---|---|---|---|

| R | 1 | 2 | 3 | 4 | 5 | 6 | 7 | 8 | Σ |
|---|---|---|---|---|---|---|---|---|---|
| R1 | 6 | 6 | 6 | 7 | 6 | 7 | 7 | 7 | |
| R2 | 4 | 4 | 5 | 5 | 5 | 6 | 6 | 7 | |
| R3 | 5 | 5 | 5 | 5 | 5 | 5 | 5 | 6 | |
| R4 | 7 | 7 | 7 | 7 | 7 | 7 | 7 | 7 | |
| R5 | 6 | 5 | 6 | 6 | 7 | 6 | 7 | 6 | |
| R6 | 4 | 4 | 4 | 4 | 4 | 6 | 6 | 7 | |
| R7 | 6 | 6 | 6 | 5 | 6 | 6 | 7 | 7 | |
| R8 | 6 | 5 | 6 | 7 | 6 | 6 | 6 | 7 | |
| R9 | 4 | 5 | 5 | 7 | 5 | 7 | 7 | 6 | |
| R10 | 6 | 6 | 6 | 6 | 6 | 6 | 6 | 7 | |
| R11 | 5 | 5 | 5 | 6 | 6 | 6 | 7 | 7 | |
| R12 | 7 | 7 | 7 | 7 | 7 | 7 | 7 | 7 | |
| R13 | 4 | 5 | 6 | 5 | 6 | 7 | 6 | 7 | |
| R14 | 1 | 2 | 2 | 1 | 2 | 3 | 4 | 4 | |
| R15 | 4 | 5 | 6 | 6 | 6 | 6 | 7 | 7 | |
| Σ | 75 | 77 | 82 | 84 | 84 | 91 | 95 | 99 | |
| Mean | 5 | 5.1 | 5.5 | 5.6 | 5.6 | 6.1 | 6.3 | 6.6 | 45.8 |
| ΣTotal | | | | | | | | | 5.7 |

**TABLE 6.** Results of SUS Assessment

| R | 1 | 2 | 3 | 4 | 5 | 6 | 7 | 8 | 9 | 10 | Σ | Σ*2.5 |
|---|---|---|---|---|---|---|---|---|---|---|---|---|
| R1 | 3 | 4 | 2 | 4 | 3 | 3 | 3 | 3 | 2 | 3 | 30 | 75 |
| R2 | 2 | 3 | 3 | 3 | 3 | 5 | 2 | 3 | 3 | 4 | 31 | 77.5 |
| R3 | 3 | 3 | 2 | 4 | 3 | 4 | 3 | 3 | 3 | 5 | 33 | 82.5 |
| R4 | 3 | 4 | 3 | 3 | 3 | 5 | 2 | 4 | 2 | 5 | 34 | 85 |
| R5 | 3 | 5 | 3 | 3 | 2 | 3 | 2 | 3 | 2 | 4 | 30 | 75 |
| R6 | 2 | 3 | 3 | 5 | 3 | 5 | 2 | 4 | 3 | 5 | 35 | 87.5 |
| R7 | 2 | 5 | 3 | 3 | 3 | 5 | 3 | 4 | 3 | 3 | 34 | 85 |
| R8 | 3 | 4 | 3 | 4 | 2 | 5 | 3 | 4 | 3 | 4 | 35 | 87.5 |
| R9 | 2 | 5 | 3 | 4 | 2 | 4 | 2 | 5 | 2 | 3 | 32 | 80 |
| R10 | 3 | 4 | 2 | 4 | 3 | 5 | 2 | 4 | 2 | 4 | 33 | 82.5 |
| R11 | 2 | 4 | 2 | 4 | 3 | 4 | 3 | 3 | 2 | 3 | 30 | 75 |
| R12 | 3 | 5 | 3 | 5 | 2 | 4 | 2 | 5 | 2 | 5 | 36 | 90 |
| R13 | 2 | 3 | 2 | 3 | 3 | 5 | 3 | 5 | 3 | 4 | 33 | 82.5 |
| R14 | 0 | 4 | 2 | 3 | 3 | 5 | 3 | 3 | 2 | 2 | 27 | 67.5 |
| R15 | 2 | 5 | 3 | 4 | 3 | 5 | 2 | 4 | 3 | 4 | 35 | 87.5 |
| Σ | | | | | | | | | | | | 1220 |
| Mean | | | | | | | | | | | | 81,3 |

## Feature Set

Figure 2 demonstrates the feature set of the dashboard system. Here, the researchers used a use case diagram for identifying the use layer diagram. There are included the five 1$^{st}$ layer features, four 2$^{nd}$ layer features, nine 3$^{rd}$ layer features, four 4$^{th}$ layer features, and around 42 last layer features

## Results of the User Experience Assessments

Table 5 presents the results of the SEQ analysis of 15 participants. Meanwhile, Table 6 shows the results of the SUS analysis. From the SEQ and SUS calculations above it can be seen that the system has an SEQ score of 5.7 and a SUS score of 81.3.

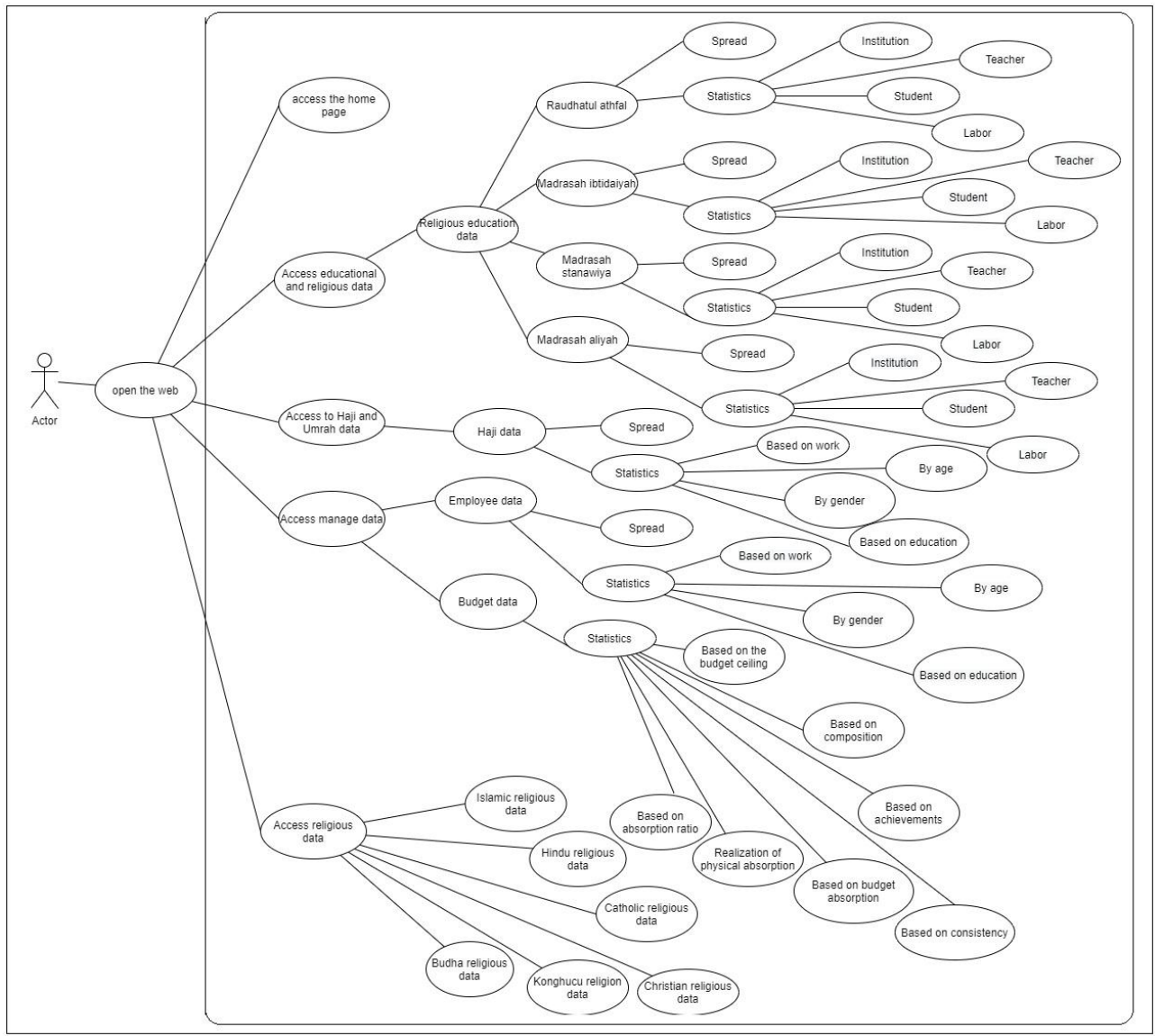

**FIGURE 2.** Feature Set of the Dashboard System

In short, we can see that results of the SEQ evaluation presented its score value around 5.7. Meanwhile, the SUS evaluation expressed the system acceptance value around 81.3 points. Referring to previous user experience studies (10-12), the SEQ analysis results indicate that the system may easy to use. Also, the SUS analysis results elucidate that the system may categorize as an acceptable system.

## CONCLUSION

In conclusion, two highlighted points of this user experience evaluation are around the easiness and eligibility aspects of the dashboard system. Both points may have answered the research question of the study. Practically, the findings may help the internal system development team, in terms of the next system development. It is related to the use of the lean product development perspective of the system. Of course, the study may present limitations issues considering the participant involvement, data collection techniques, or the subjectivity of the researchers involved in the study. Thus, the above-mentioned issues may be one of the considerations for the next similar studies.